\documentclass{article}

\usepackage{amsmath}
\usepackage{amssymb}
\usepackage{multirow}

\begin{document}

\title{Two-player conflicting interest Bayesian games and Bell nonlocality\thanks{The final publication is available at Springer via http://dx.doi.org/10.1007/s11128-015-1171-1}}

\author{Haozhen~Situ
\thanks{H.Z. Situ is with the College of Mathematics and Informatics, South China Agricultural University, Guangzhou,
 510642 China. E-mail: situhaozhen@gmail.com}}

\maketitle

\begin{abstract}
Nonlocality, one of the most remarkable aspects of quantum mechanics, is closely related
to Bayesian game theory.
Quantum mechanics can offer advantages to some Bayesian games,
if the payoff functions are related to Bell inequalities in some way.
Most of these Bayesian games that have been discussed are common interest games.
Recently the first conflicting interest Bayesian game is proposed in Phys. Rev. Lett. 114, 020401 (2015).
In the present paper we present three new conflicting interest Bayesian games where quantum mechanics offers advantages.
The first game is linked with Cereceda inequalities,
the second game is linked with a generalized Bell inequality with 3 possible measurement outcomes,
and the third game is linked with a generalized Bell inequality with 3 possible measurement settings.
\end{abstract}

\noindent\textbf{Keywords}:
Bayesian game, nonlocal game, Bell inequalities.\\\\

\section{Introduction}

Nonlocality is one of the most remarkable aspects of quantum mechanics.
It can generate quantum correlation between two remote observers which are stronger than classical correlation explained by local hidden variables.
The readers can refer to a recent review article \cite{BCPSW} for a general overview of nonlocality.
Interestingly, it happens that this physics concept has something in common with a concept of game theory, while quantum mechanics and game theory are two seemingly disparate subjects.
It has been pointed out that Bell inequalities \cite{Bell} and Bayesian games \cite{Harsanyi} are closely related \cite{CI08,BL13,ICA15}.
The Bell expression and the expected payoff funciton have the same form so that
if quantum mechanics breaks one specific Bell inequality, it can generate nonlocal correlation between the players and gives the corresponding expected payoff function a higher value.
Bayesian games that are related to Bell inequalities are often called ``nonlocal games'' in the literature.
The GHZ-Mermin game \cite{GHZ}, the Magic Square game \cite{magicsquare1,magicsquare2}, the Hidden Matching game \cite{hiddenmatching1,hiddenmatching2} are all examples of nonlocal games.
In these nonlocal games the players have common interests.

Recently, Pappa \emph{et al.} \cite{Pappa15} present the first conflicting interest Bayesian game in which quantum mechanics leads to a higher total payoff, and they suggest that it would be interesting to find more conflicting interest games where quantum mechanics offers an advantage.
In the present paper we present three new such games.
The first game is linked with Cereceda inequalities \cite{Cereceda},
the second game is linked with a generalized Bell inequality with 3 possible measurement outcomes \cite{Collins},
and the third game is linked with a generalized Bell inequality with 3 possible measurement settings \cite{Wehner}.

\section{A Bayesian Game Linked with Cereceda Inequality}

Let's begin by describing our first game between two players.
When the game starts, each player will receive a random binary input.
We use the notation $A_1,A_2$ to denote the two possible values of Alice's input and $B_1,B_2$ for Bob's. Each player knows her/his own input but not the other player's input.
(In other words, the players have partial information about the setting in which the game is played.)
Each player then has to decide a binary output alone.
We use $a_i\in\{0,1\}$ to denote Alice's output when her input is $A_i$, and $b_i\in\{0,1\}$ for Bob's output.
So we can regard $a_1a_2$ as Alice's strategy, and $b_1b_2$ Bob's.
The payoffs for them are determined by the inputs and the outputs.
The mapping can be described by payoff functions or payoff matrices.
The payoff matrix of our first game is given in Table \ref{tab:game1matrix}.
The two inputs $A_iB_j$ fix a $2\times 2$ block and the two outputs $a_ib_j$'s four possible values $00,01,10,11$ corresponds to the four entries (from left to right, up to down) of the block.
Each entry is a tuple of two values, the first value for Alice's payoff and the second one for Bob's.
For example, if Alice receives $A_1$ and outputs 1, Bob receives $B_2$ and outputs 0,
then Alice obtains payoff 1 and Bob obtains 2.

\begin{table}
\caption{Payoff matrix of our first game.}
\label{tab:game1matrix}
\centering
\begin{tabular}{c|cc|cc}
& \multicolumn{2}{|c|}{$B_1$}  & \multicolumn{2}{|c}{$B_2$}\\
\hline
\multirow{2}{*}{$A_1$} & (2,1) & (0,0) & (0,0) & (2,1)\\
& (0,0) & (1,2) & (1,2) & (0,0)\\
\hline
\multirow{2}{*}{$A_2$} & (0,0) & (1,2) & $(-\frac{3}{2},-\frac{3}{2})$ & (0,0)\\
& (2,1) & (0,0) & (0,0) & ($-\frac{3}{2}$,$-\frac{3}{2}$) \\
\end{tabular}
\end{table}

Because the inputs are random variables, we have to consider the expected payoffs.
The expected payoff functions for Alice and Bob are
\begin{align}
\$_A(a_1a_2b_1b_2) & = Pr(A_1B_1) \big(2\cdot P^{11}_{00}+ 1\cdot P^{11}_{11}\big)
 +Pr(A_1B_2) \big(2\cdot P^{12}_{01}+ 1\cdot P^{12}_{10}\big)\nonumber\\
& +Pr(A_2B_1) \big(1\cdot P^{21}_{01}+ 2\cdot P^{21}_{10}\big)
 +Pr(A_2B_2) \big(-\frac{3}{2}\cdot P^{22}_{00} -\frac{3}{2}\cdot P^{22}_{11}\big),\\
\$_B(a_1a_2b_1b_2) & = Pr(A_1B_1) \big(1\cdot P^{11}_{00}+ 2\cdot P^{11}_{11}\big)
 +Pr(A_1B_2) \big(1\cdot P^{12}_{01}+ 2\cdot P^{12}_{10}\big)\nonumber\\
& +Pr(A_2B_1) \big(2\cdot P^{21}_{01}+ 1\cdot P^{21}_{10}\big)
 +Pr(A_2B_2) \big(-\frac{3}{2}\cdot P^{22}_{00} -\frac{3}{2}\cdot P^{22}_{11}\big),
\end{align}
where
\begin{align}
P^{ij}_{kl}=\begin{cases}
1,   & a_i=k \ and\  b_j=l \\
0,   & otherwise.
\end{cases}
\end{align}
We suppose the inputs are uniformly random and set $Pr(A_iB_j)=\frac{1}{4}$ for all $i,j$.
We list all the 16 combinations of Alice's and Bob's possible outputs and the corresponding expected payoffs in Table \ref{tab:game1equilibrium}.

\begin{table}
\caption{Expected payoffs of our first game.}
\label{tab:game1equilibrium}
\centering
\begin{tabular}{cccc|ccc|c}
$a_1$ & $a_2$ & $b_1$ & $b_2$ & $\$_A$ & $\$_B$ & $\$_A+\$_B$ & Equilibrium\\
\hline
0 & 0 & 0 & 0 & 1/8 & -1/8 & 0 & no\\
0 & 0 & 0 & 1 & 1 & 1/2 & 3/2 & no\\
0 & 0 & 1 & 0 & -1/8 & 1/8 & 0 &no\\
0 & 0 & 1 & 1 & 3/4 & 3/4 & 3/2 & YES\\
0 & 1 & 0 & 0 & 1 & 1/2 & 3/2 & YES\\
0 & 1 & 0 & 1 & 9/8 & 3/8 & 3/2 & no\\
0 & 1 & 1 & 0 & 0 & 0 & 0 & no\\
0 & 1 & 1 & 1 & 1/8 & -1/8 & 0 & no\\
1 & 0 & 0 & 0 & -1/8 & 1/8 & 0 & no\\
1 & 0 & 0 & 1 & 0 & 0 & 0 & no\\
1 & 0 & 1 & 0 & 3/8 & 9/8 & 3/2 & no\\
1 & 0 & 1 & 1 & 1/2 & 1 & 3/2 & no\\
1 & 1 & 0 & 0 & 3/4 & 3/4 & 3/2 & no\\
1 & 1 & 0 & 1 & 1/8 & -1/8 & 0 & no\\
1 & 1 & 1 & 0 & 1/2 & 1 & 3/2 & YES\\
1 & 1 & 1 & 1 & -1/8 & 1/8 & 0 & no\\
\end{tabular}
\end{table}

From Table \ref{tab:game1equilibrium} we can see that there are three equilibria.
Equilibrium is a key concept in game theory, which describes a stable state in which each player's output is the best response to the other player(s)' output(s).
For example, $\$_A(0011)$ is the largest of $\{\$_A(a_1a_211)|\forall a_1,a_2\}$ and $\$_B(0011)$ is the largest of $\{\$_B(00b_1b_2)|\forall b_1,b_2\}$ so the output 0011 is an equilibrium point.
It's a fair equilibrium because $\$_A(0011)=\$_B(0011)$.
However, Alice prefers the equilibrium 0100 with payoff $(1,\frac{1}{2})$ while Bob prefers 1110 with payoff $(\frac{1}{2},1)$ because game theory assumes the players are selfish and aim to maximize their own profits.
This game is a conflicting interest game because the two players cannot agree on the same equilibrium.
From the table we can also see that the largest total payoff $\$_A+\$_B$ is $3/2$, even if randomness is incorporated in the decision process.

Randomness can be incorporated in the form of common advice.
A probability distribution on 16 output combinations has to be generated by an arbitrator before the game starts. (So the distribution doesn't depend on the inputs because the inputs are fed to the players after the game has started.)
After the game has started one combination $a_1a_2b_1b_2$ is sampled from the distribution by the arbitrator and then delivered to the two players.
Alice/Bob outputs $a_i/b_j$ if her/his input is $A_i/B_j$.
With the help of advice their outputs may be correlated.
For instance, the distribution $Pr(a_1a_2b_1b_2=0000)=Pr(a_1a_2b_1b_2=1111)=\frac{1}{2}$ makes the players output the same random value,
while the distribution $Pr(a_1a_2b_1b_2=0011)=Pr(a_1a_2b_1b_2=1100)=\frac{1}{2}$ makes the players output two different random values.
In the case of playing the game with randomness, the expected payoffs for Alice and Bob are
\begin{align}\label{eq:Apayoff}
\$_A & =\sum_{i,j,k,l=0}^1 Pr(a_1a_2b_1b_2=ijkl)\cdot \$_A(ijkl)\nonumber\\
& = \frac{1}{4}\big(2 P^{11}_{00}+  P^{11}_{11}
+ 2 P^{12}_{01}+  P^{12}_{10} +  P^{21}_{01}+ 2 P^{21}_{10}
-\frac{3}{2} P^{22}_{00} -\frac{3}{2} P^{22}_{11}\big),\\
\label{eq:Bpayoff}
\$_B & =\sum_{i,j,k,l=0}^1 Pr(a_1a_2b_1b_2=ijkl)\cdot \$_B(ijkl)\nonumber\\
& = \frac{1}{4}\big( P^{11}_{00}+ 2 P^{11}_{11}
+  P^{12}_{01}+ 2 P^{12}_{10} +  2 P^{21}_{01}+  P^{21}_{10}
-\frac{3}{2} P^{22}_{00} -\frac{3}{2} P^{22}_{11}\big),
\end{align}
where
\begin{align}\label{eq:classicalprobabilities}
P^{ij}_{kl}=\sum_{m,n=0}^1 Pr(a_i=k,b_j=l,a_{3-i}=m,b_{3-j}=n)
\end{align}
represents the probability of outputing $kl$ given the inputs $A_iB_j$.

If we compare the above scenario with the Bell experiment \cite{Bell},
we can find that $A_1,A_2$ can be regarded as Alice's two types of measurements,
$a_1,a_2$ as the measurement outcomes, and analogously for $B_1,B_2$ and $b_1,b_2$,
and then Eq. (\ref{eq:Apayoff}) and (\ref{eq:Bpayoff}) are actually Bell expressions.
More interestingly, the total payoff
\begin{align}
\$_A+\$_B=\frac{3}{4}(P^{11}_{11}+P^{12}_{01}+P^{21}_{10}-P^{22}_{11})+
\frac{3}{4}(P^{11}_{00}+P^{12}_{10}+P^{21}_{01}-P^{22}_{00})
\end{align}
is propotional to the left part of the sum of the following two Cereceda inequalities \cite{Cereceda}:
\begin{align}
P^{11}_{11}+P^{12}_{01}+P^{21}_{10}-P^{22}_{11}\leqslant 1,\\
P^{11}_{00}+P^{12}_{10}+P^{21}_{01}-P^{22}_{00}\leqslant 1.
\end{align}
The upper bound 1 is induced by local realism.
Thus $\$_A+\$_B\leqslant\frac{3}{2}$, congruent with the largest total payoff in Table. \ref{tab:game1equilibrium}.

However, probabilities generated by quantum mechanics can violate these inequalities.
An appropriate choice of observables $A_1,A_2,B_1,B_2$ and a maximally entangled state of two qubits can achieve
\begin{align}
P^{11}_{11}+P^{12}_{01}+P^{21}_{10}-P^{22}_{11}=\frac{1+\sqrt{2}}{2}>1,\\
P^{11}_{00}+P^{12}_{10}+P^{21}_{01}-P^{22}_{00}=\frac{1+\sqrt{2}}{2}>1.
\end{align}
Here $P^{ij}_{kl}$ are probabilities generated by quantum resource, not the same as
the probabilities generated by classical resource (Eq. (\ref{eq:classicalprobabilities})).
Suppose the quantum state is $|\phi\rangle$,
The eigenvectors of $A_1$ and $A_2$ are $\{|\alpha^1_0\rangle\langle\alpha^1_0|,|\alpha^1_1\rangle\langle\alpha^1_1|\}$
and
$\{|\alpha^2_0\rangle\langle\alpha^2_0|,|\alpha^2_1\rangle\langle\alpha^2_1|\}$.
The eigenvectors of $B_1$ and $B_2$ are
$\{|\beta^1_0\rangle\langle\beta^1_0|,|\beta^1_1\rangle\langle\beta^1_1|\}$
and
$\{|\beta^2_0\rangle\langle\beta^2_0|,|\beta^2_1\rangle\langle\beta^2_1|\}$.
The probabilities can be given by
\begin{align}
P^{ij}_{kl}=|\langle\alpha^i_k,\beta^j_l| \phi\rangle|^2.
\end{align}
$\frac{1+\sqrt{2}}{2}$ is the maximal violation allowed by quantum mechanics \cite{Tsirelson}.
So players with quantum resource can obtain $\$_A+\$_B=\frac{3}{4}(1+\sqrt{2})>\frac{3}{2}$.
At the same time, $\$_A=\$_B=\frac{3}{8}(1+\sqrt{2})$,
a fair allocation is achieved.

\section{A Bayesian Game with Ternary Outputs}

In Ref. \cite{Pappa15} the authors suggest that
it would be interesting to find more conflicting interest games
where quantum mechanics offers an advantage,
for example when larger dimensions are used.
In this section we explore the relation between a 3 dimensional Bell inequality and a Bayesian game.
In this case the outputs of the Bayesian game are ternary rather than binary, i.e.,
$a_i,b_i\in\{0,1,2\}$ for $i=1,2$.
The payoff matrix of our second game is given in Table \ref{tab:game2matrix}.
It is a conflicting interest game,
because Alice prefers equilibrium $a_1a_2b_1b_2=0010$ with payoff $(\frac{11}{8},\frac{7}{8})$
while Bob prefers equilibrium $a_1a_2b_1b_2=2122$ with payoff $(\frac{7}{8},\frac{11}{8})$.

\begin{table}
\caption{Payoff matrix of our second game.}
\label{tab:game2matrix}
\centering
\begin{tabular}{c|ccc|ccc}
& \multicolumn{3}{|c|}{$B_1$}  & \multicolumn{3}{|c}{$B_2$}\\
\hline
\multirow{3}{*}{$A_1$} & (2,1) & (0,0) & (0,0) & (2,1) & (0,0) & (0,0)\\
& (0,0) & ($\frac{3}{2}$,$\frac{3}{2}$) & (0,0) & (0,0) & ($\frac{3}{2}$,$\frac{3}{2}$) & (0,0)\\
& (0,0) & (0,0) & (1,2) & (0,0) & (0,0) & (1,2)\\
\hline
\multirow{3}{*}{$A_2$} & (0,0) & ($\frac{3}{2}$,$\frac{3}{2}$) & (0,0) & (2,1) & (0,0) & (0,0)\\
& (0,0) & (0,0) & ($\frac{3}{2}$,$\frac{3}{2}$) & (0,0) & ($\frac{3}{2}$,$\frac{3}{2}$) & (0,0) \\
& ($\frac{3}{2}$,$\frac{3}{2}$) & (0,0) & (0,0) & (0,0) & (0,0) & (1,2)
\end{tabular}
\end{table}

The expected payoffs for Alice and Bob are
\begin{align}
\$_A=&\frac{1}{4}(2\cdot P^{11}_{00}+\frac{3}{2}\cdot P^{11}_{11}+P^{11}_{22}
+2\cdot P^{12}_{00}+\frac{3}{2}\cdot P^{12}_{11}+P^{12}_{22}\nonumber\\
&+\frac{3}{2}\cdot P^{21}_{01}+\frac{3}{2}\cdot P^{21}_{12}+\frac{3}{2}\cdot P^{21}_{20}
+2\cdot P^{22}_{00}+\frac{3}{2}\cdot P^{22}_{11}+ P^{22}_{22}),\\
\$_B=&\frac{1}{4}(P^{11}_{00}+\frac{3}{2}\cdot P^{11}_{11}+2\cdot P^{11}_{22}
+P^{12}_{00}+\frac{3}{2}\cdot P^{12}_{11}+2\cdot P^{12}_{22}\nonumber\\
&+\frac{3}{2}\cdot P^{21}_{01}+\frac{3}{2}\cdot P^{21}_{12}+\frac{3}{2}\cdot P^{21}_{20}
+ P^{22}_{00}+\frac{3}{2}\cdot P^{22}_{11}+ 2\cdot P^{22}_{22}).
\end{align}
The total payoff $\$_A+\$_B$ is $\frac{3}{4} C$ with
\begin{align}
C=& P^{11}_{00}+ P^{11}_{11}+P^{11}_{22}
+ P^{12}_{00}+ P^{12}_{11}+P^{12}_{22}\nonumber\\
&+ P^{21}_{01}+P^{21}_{12}+ P^{21}_{20}
+ P^{22}_{00}+ P^{22}_{11}+ P^{22}_{22},
\end{align}
which is the left part of a generalized Bell inequality $C\leqslant 3$ \cite{Collins}.
Thus $\$_A+\$_B\leqslant\frac{9}{4}$.

By choosing appropriate measurements on a maximally entangled state of two qutrits, the players can achieve
\begin{align}
C=\frac{2}{9 \sin^2(\frac{\pi}{12})}>3.
\end{align}
So players with quantum resource can obtain $\$_A+\$_B=\frac{3}{4}C=(6 \sin^2(\pi/12))^{-1}>\frac{9}{4}$.
At the same time, $\$_A=\$_B=(12 \sin^2(\pi/12))^{-1}$,
a fair allocation is achieved.

\section{A Bayesian Game with Ternary Inputs}

In this section we present a Bayesian game with ternary inputs.
The three possible values of Alice's input are denoted as $A_1,A_2,A_3$,
and the corresponding outputs for each input value are denoted as $a_1,a_2,a_3$.
Analogously, the notations for Bob's input values and outputs are $B_1,B_2,B_3$ and $b_1,b_2,b_3$.
The payoff matrix of our third game is given in Table \ref{tab:game3matrix}.
It is a conflicting interest game,
because Alice prefers equilibrium $a_1a_2a_3b_1b_2b_3=000001$ with payoff $(\frac{5}{9},\frac{1}{9})$
while Bob prefers equilibrium $a_1a_2a_3b_1b_2b_3=011111$ with payoff $(\frac{1}{9},\frac{5}{9})$.

\begin{table}
\caption{Payoff matrix of our third game.}
\label{tab:game3matrix}
\centering
\begin{tabular}{c|cc|cc|cc}
& \multicolumn{2}{|c|}{$B_1$}  & \multicolumn{2}{|c}{$B_2$} & \multicolumn{2}{|c}{$B_3$}\\
\hline
\multirow{2}{*}{$A_1$} & (2,1) & (0,0) & (-$\frac{3}{2}$,-$\frac{3}{2}$) & (-$\frac{3}{2}$,-$\frac{3}{2}$) & (0,0) & ($\frac{3}{2}$,$\frac{3}{2}$)\\
& (0,0) & (1,2) & (-$\frac{3}{2}$,-$\frac{3}{2}$) & (-$\frac{3}{2}$,-$\frac{3}{2}$) & ($\frac{3}{2}$,$\frac{3}{2}$) & (0,0)\\
\hline
\multirow{2}{*}{$A_2$} & (2,1) & (0,0) & (2,1) & (0,0) & (-$\frac{3}{2}$,-$\frac{3}{2}$) & (-$\frac{3}{2}$,-$\frac{3}{2}$)\\
 & (0,0) & (1,2) & (0,0) & (1,2) & (-$\frac{3}{2}$,-$\frac{3}{2}$) & (-$\frac{3}{2}$,-$\frac{3}{2}$)\\
\hline
\multirow{2}{*}{$A_3$} & (-$\frac{3}{2}$,-$\frac{3}{2}$) & (-$\frac{3}{2}$,-$\frac{3}{2}$) & (2,1) & (0,0) & (2,1) & (0,0) \\
& (-$\frac{3}{2}$,-$\frac{3}{2}$) & (-$\frac{3}{2}$,-$\frac{3}{2}$) & (0,0) & (1,2) & (0,0) & (1,2)
\end{tabular}
\end{table}

The expected payoffs for Alice and Bob are
\begin{align}
\$_A=&\frac{1}{9}(2\cdot P^{11}_{00}+P^{11}_{11}+2\cdot P^{21}_{00}+P^{21}_{11}+2\cdot P^{22}_{00}+P^{22}_{11}
+2\cdot P^{32}_{00}+P^{32}_{11}\nonumber\\
& +2\cdot P^{33}_{00}+P^{33}_{11}
+\frac{3}{2}P^{13}_{01}+\frac{3}{2}P^{13}_{10}-\frac{3}{2}\cdot 3),\\
\$_B=&\frac{1}{9}(P^{11}_{00}+2\cdot P^{11}_{11}+ P^{21}_{00}+2\cdot P^{21}_{11}+ P^{22}_{00}+2\cdot P^{22}_{11}
+ P^{32}_{00}+2\cdot P^{32}_{11}\nonumber\\
& + P^{33}_{00}+2\cdot P^{33}_{11}
+\frac{3}{2}P^{13}_{01}+\frac{3}{2}P^{13}_{10}-\frac{3}{2}\cdot 3).
\end{align}
The total payoff $\$_A+\$_B$ is $\frac{1}{6} C$ with
\begin{align}
C=& \langle A_1B_1\rangle +\langle A_2B_2\rangle+\langle A_3B_3\rangle
+\langle A_2B_1\rangle+\langle A_3B_2\rangle-\langle A_1B_3\rangle,
\end{align}
where
\begin{align}
\langle A_iB_j\rangle=P^{ij}_{00}+ P^{ij}_{11} -P^{ij}_{01}-P^{ij}_{10}
=2(P^{ij}_{00}+ P^{ij}_{11})-1.
\end{align}
$C$ is the left part of a generalized Bell inequality $C\leqslant 4$ \cite{Wehner}.
Thus $\$_A+\$_B\leqslant\frac{2}{3}$.

By choosing appropriate measurements on a maximally entangled state of two qubits, the players can achieve
\begin{align}
C=6 \cos\frac{\pi}{6}>4.
\end{align}
So players with quantum resource can obtain $\$_A+\$_B=\frac{1}{6} C=\cos \frac{\pi}{6}>\frac{2}{3}$.
At the same time, $\$_A=\$_B=\frac{1}{2}\cos \frac{\pi}{6}$,
a fair allocation is achieved.

\section{Conclusion}

In this paper we have presented three new conflicting interest Bayesian games where quantum mechanics offers advantages.
The first game is linked with Cereceda inequalities,
the second game is linked with a generalized Bell inequality with 3 possible measurement outcomes,
and the third game is linked with a generalized Bell inequality with 3 possible measurement settings.
The game presented in Ref. \cite{CI08} is also linked with Cereceda inequalities,
but it's not a conflicting interest game.
For that game each player's payoff function corresponds to a linear combination of Cereceda inequalities.
However, for our first game neither player's payoff function corresponds to a linear combination of Cereceda inequalities but the sum of their payoff functions does.
Similarly, for the first conflicting game presented in Ref. \cite{Pappa15},
neither player's payoff function corresponds to a linear combination of CHSH inequalities \cite{CHSH},
but the sum of their payoff functions corresponds to a CHSH inequality.
This may explain the rise of conflicting interests.
This paper may contribute to our understanding of both Bell nonlocality and Bayesian game theory,
and may be useful in further study on relevant topics.

\noindent \textbf{Acknowledgements}
This work is supported by
the National Natural Science Foundation of China (Grant Nos. 61502179, 61472452)
and
the Natural Science Foundation of Guangdong Province of China (Grant No. 2014A030310265).

\end{document}